\documentclass[prd,aps,preprint,epsfig,floats,superscriptaddress]{revtex4}

\usepackage{graphicx}
\usepackage{times}
\usepackage{epsfig}
\usepackage{amsmath}
\usepackage{amsfonts}
\usepackage{amssymb}
\usepackage{url}
\usepackage{hyperref}
\newcommand{\be}{\begin{equation}}
\newcommand{\ee}{\end{equation}}
\newcommand{\bea}{\begin{eqnarray}}
\newcommand{\eea}{\end{eqnarray}}

\begin{document}

\pagestyle{plain}

\title{A Model With Dynamical R-parity Breaking \\ and Unstable
Gravitino Dark Matter}

\author{Xiangdong Ji}
 \affiliation{Maryland Center for Fundamental Physics and Department of Physics, University of
Maryland, College Park, Maryland 20742, USA } \affiliation{Center
for High-Energy Physics and Institute of Theoretical Physics, Peking
University, Beijing 100871, China}
\author{Rabindra N. Mohapatra}
 \affiliation{Maryland Center for Fundamental Physics and Department of Physics, University of
Maryland, College Park, Maryland 20742, USA }
\author{Shmuel Nussinov}
\affiliation{Tel Aviv University, Israel}
\author{Yue Zhang}
\affiliation{Center for High-Energy Physics and Institute of
Theoretical Physics, Peking University, Beijing 100871, China}
 \affiliation{Maryland Center for Fundamental Physics and Department of
Physics, University of Maryland, College Park, Maryland 20742, USA }

\date{\today}

\preprint{\vbox{\hbox{UMD-PP-08-013}}}

\begin{abstract}
An unstable gravitino with lifetime longer than $10^{26}$ sec or so
has been proposed as a possible dark matter candidate in supergravity
models with R-parity breaking. We find a natural realization of this
idea in the minimal supersymmetric left-right models where
left-right symmetry breaking scale in the
few TeV range. It is known that in these models, R-parity must break
in order to have parity breaking as required by low energy weak 
interactions. The sub-eV neutrino masses imply that R-parity breaking 
effects in this model must be highly suppressed. This in turn 
makes the gravitino LSP long lived enough, so that it becomes the dark 
matter of
the Universe. It also allows detectable displaced vertices at the LHC 
from NLSP decays. We
present a detailed analysis of the model and some aspects of its rich
phenomenology.
\end{abstract}
\maketitle 

\section{Introduction} \ \
It is now widely believed that about 25\%
of the energy density in the Universe is in the form of a cold dark
matter. The nature of the elementary particle which constitutes this dark
component is however not known. The scenario with an unstable gravitino
which is the lightest supersymmetric particle (LSP) with a very long life time
as a dark matter is particularly attractive~\cite{gravilsp,gravilsp1},
having interesting consequences for astrophysics and cosmology, as well
as colliders.

The LSP gravitino in an R-parity conserving supersymmetric (SUSY) theory
is absolutely stable and could be a dark matter
candidate~\cite{feng}. An unstable
gravitino, in a theory with broken R-parity symmetry, needs to be sufficiently
long lived to be a dark matter. The strengths of the R-parity violating
(RPV) couplings (usually denoted by $\lambda,\lambda'$) needed to achive
required longevity must be highly
suppressed i.e. $\lambda,\lambda'\leq 10^{-6}$, and one would
then like to understand the origin of such small couplings.
It is therefore interesting to explore models where such
small couplings may arise naturally. We find
that TeV scale supersymmetric left-right (SUSYLR) models which have been
discussed in connection with neutrino
masses and strong CP problems provides one such framework, with two 
interesting features: (a) R-parity breaking is dynamically induced in 
the global minimum of the potential in order for the theory to break 
parity; (b) smallness of neutrino masses guarantees that the resulting 
R-parity violating interactions are highly suppressed.

The left-right symmetric theories based on the gauge group $SU(2)_L\times 
SU(2)_R\times U(1)_{B-L}$\cite{LR} were originally introduced to 
explain parity violation in the standard model and were 
found to have a number of interesting properties. These models provide a 
natural framework for understanding the small neutrino masses via the 
seesaw mechanism\cite{seesaw} and became specially 
interesting after the discovery of neutrino masses. The 
 right handed neutrino which is essential for seesaw mechanism is 
automatically contained
and furthermore the B-L gauge symmetry\cite{marshak} whose breaking 
provides the heavy Majorana mass to the right handed neutrinos relates 
the small neutrino masses to the parity breaking scale. This 
implementation of the seesaw mechanism is different from those in the SM 
context where the right handed neutrino mass is a free 
parameter. 

The seesaw scale (or the parity breaking scale) however, 
still remains undetermined. It can take values anywhere from
near a TeV  if the Dirac Yukawa couplings $Y_{\nu}\sim
10^{-6}\approx Y_e$ or it can be much higher if the $Y_\nu$'s are larger. 
It is important to note that a TeV seesaw scale is perfectly natural and 
does not require any higher fine tuning than that present in SM. A TeV 
seesaw scale is clearly of great interest for the LHC. It is worth noting 
that the current searches by the CDF and D0
collaborations have yielded limits on the parity breaking scale in the 
750 GeV range~\cite{fermilab}.



In this paper we consider the minimal left-right (SUSYLR) 
extension of the MSSM since it combines the advantages of supersymmetry 
while making MSSM realistic by providing a way to understand neutrino 
masses. It also cures certain problems of the MSSM such as making
R-parity a good symmetry if one uses B-L=2 Higgs triplets to break 
parity so that proton is stable. Furthermore, it provides a solution to 
the SUSY and strong CP problems. The minimal version of this model 
(without any gauge singlets) has two striking 
features: (i) the parity and SUSY breaking scales are related, thereby 
predicting that the seesaw scale is necessarily in the TeV range and
(ii) while R-parity conservation is automatic above the parity
breaking scale, the ground state of the theory can break parity
only if R-parity is spontaneously broken by a vaccum expectation value 
(VEV) of the right-handed sneutrino fields~\cite{kuchi}. Thus 
the model breaks R-parity dynamically. Furthermore in 
this case there is an
upper limit on the $W_R$ scale in the range of a few 
TeV's~\cite{kuchi1}, which is to be expected since parity breaking and 
susy breaking are intimately linked. This makes it possible to test
the theory using LHC data expected in coming years. 

An interesting 
feature of the resulting dynamical R-parity breaking is that it only 
breaks lepton number and keeps baryon
number intact and therefore the proton is absolutely stable in this 
model. It is also worth noting that while the
effective R-parity breaking below the $W_R$ scale has some properties 
similar to MSSM with bilinear R-parity breaking~\cite{valle}, it has many 
properties which are characteristic of the SUSYLR theory that can 
provide distinguishing tests.

The immediate question that then arises is whether this SUSY model which
starts out promising a stable dark matter, does indeed have a dark
matter after parity and R-parity breaking. We
address this question in this paper.
We find that despite R-parity breaking, the unstable LSP gravitino in our 
model can be the dark matter of the
universe~\cite{gravilsp}. The reason for this is that requiring
sub-eV neutrino masses suppresses the strengths of the R-parity breaking
interactions responsible for gravitino decay to such a level that 
 the gravitino becomes sufficiently long lived and dark matter.  
Secondly, we also find that despite the neutrino-Higgsino mixing
induced by spontaneous R-parity breaking, the seesaw results for
neutrino masses remain essentially intact. Another consequence of
this small strength of R-parity breaking is that the next-to-lightest 
SUSY particle (NLSP) which can be a
neutralino or stau or sneutrino produced at LHC has a
lifetime such that it can give rise to displaced vertices in the LHC
detector~\cite{bar}.

The paper is organized as follows: Section II presents the basic
contents of the model, and reviews the result of Ref.~\cite{kuchi}
that R-parity indeed must break in the model if ordinary parity has to 
break. In Section III, we consider implications of RPV on the
neutrino mass in this model.
In Section IV, we discuss contribution to various RPV couplings
due to bilinear terms in the superpotential.
In Section V, we derive cosmological implications for gravitino as
the dark matter. In Section VI, we
discuss collider signatures of the model and in Section VII, we briefly discuss
some other consequences of the model. In the appendix,
we display the
minimization of the potential and obtain bounds on the $W_R$-boson mass and
the right-handed sneutrino VEV.

\section{Basic features of SUSYLR model and spontaneous R-parity violation}

The gauge group in the minimal SUSYLR model is $SU(2)_L\times
SU(2)_R\times U(1)_{B-L}\times SU(3)_c$. The chiral left and right handed
quark superfields are $Q\equiv (u,d)(2,1,\frac{1}{3}, 3)$ and $Q^c\equiv
(u^c, d^c)(1,2,-\frac{1}{3}, 3^*)$ respectively,
and similarly the lepton superfields are given by $L\equiv (\nu,
e)(2,1,-1,1)$ and $L^c\equiv (\nu^c, e^c)(2,1,+1,1)$, where flavor  
indices have been
implicit.
  The symmetry breaking is achieved by
the following bi-fundamental and $B-L=\pm 2 $ triplet Higgs superfields:
$\phi_a(2,2,0,1)$ ($a=1,2$),
$\Delta (3, 1, +2, 1)$, $\bar{\Delta}(3,1,-2,1)$,  $\Delta^c (3, 1, -2, 
1)$, $\bar{\Delta}^c(3,1,+2,1)$. The superpotential
of the model is:
\begin{eqnarray}
 W &=& Y_u Q^T \tau_2 \Phi_1 \tau_2 Q^c +  Y_d Q^T \tau_2 \Phi_2 \tau_2
Q^c \nonumber \\
 &+& Y_\nu L^T \tau_2 \Phi_1 \tau_2 L^c  + Y_l L^T \tau_2 \Phi_2 \tau_2
L^c \nonumber \\
 &+& i f \left( L^T \tau_2 \Delta L + L^{cT} \tau_2 \Delta^c L^c
\right) + \mu_{ab} {\rm Tr} \left( \Phi_a^T \tau_2 \Phi_b \tau_2 \right)
\nonumber \\
 &+&
 \mu_\Delta {\rm Tr} \left( \Delta \bar \Delta + \Delta^c \bar \Delta^c
\right) \ ,
\end{eqnarray}
where $Y$'s are Yukawa couplings, $f$ is the Majorana coupling and 
$\mu_\Delta$ is the $\mu$-term for triplets. Note that we do not have any 
gauge singlet fields in the model.

First point to note is that if $M_R\gg M_{SUSY}$, the SUSY breaking 
scale, the right handed gauge
symmetry remains unbroken since we must have
$\langle\Delta^c\rangle=\langle\bar{\Delta}^c\rangle=0$ to preserve 
supersymmetry. From this it
follows that these vevs i.e. $\langle\Delta^c\rangle=v_R$ and
$\langle\bar{\Delta}^c\rangle=\bar{v}_R$ and $\mu$-term must have TeV 
scale vevs, i.e., the theory is
necessarily one with TeV scale parity breaking. From now on we will
assume that all the mass parameters in the theory are of order of a TeV.

To do a detailed analysis of the ground state of the theory, we write
down the scalar
potential including the soft SUSY-breaking terms:
\begin{eqnarray}
V~=~V_F~+~V_D+~V_S \ ,
\end{eqnarray}
We first give $V_D$:
\begin{eqnarray}
 V_D &=& \frac{g_R^2}{8} \sum_m \left| \tilde \nu^{c\dag} \tilde \nu^c 
\delta_{m3} +
2 {\rm Tr} \left( \Delta^{c\dag} \tau_m \Delta^c + \bar \Delta^{c\dag}
\tau_m \bar \Delta^c \right)+ {\rm Tr} (\Phi\tau^T_m\Phi^\dagger) \right|^2
\nonumber \\
& +& \frac{g_L^2}{8} \sum_m \left| 2 {\rm Tr} \left( \Delta^{\dag}
\tau_m  \Delta + \bar \Delta^{\dag} \tau_m \bar \Delta
\right)+ {\rm Tr} (\Phi^\dagger\tau_m\Phi) \right|^2
\nonumber \\
&+&  \frac{g^{2}_{BL}}{8} \left| \tilde \nu^{c\dag} \tilde \nu^c + 2 {\rm Tr}
 \left( \Delta^{\dag} \Delta - \Delta^{c\dag} \Delta^c + \bar
 \Delta^{\dag} \bar \Delta - \bar \Delta^{c\dag} \bar \Delta^c \right)
\right|^2 \ .
\end{eqnarray}
Below we give some of the terms from $V_F =\sum_a |\frac{\partial 
W}{\partial \phi_a}|^2$, with
$\phi_a$ going over all the fields in the model and the soft susy 
breaking term $V_S$ \cite{kuchi,huitu} which are relevant for our 
discussion:
\begin{eqnarray}
V_F~+~V_S &=&
 \left[ A_L^a \widetilde  L^T \tau_2 \Phi_a \tau_2 \widetilde  L^c \right.  \\
 &+&  i A
 \left( \widetilde  L^T \tau_2 \Delta \widetilde  L + \widetilde  L^{cT}
\tau_2 \Delta^c \widetilde  L^c \right) \nonumber \\
 &+& \left. b_{ab} {\rm Tr} \left( \Phi_a^T \tau_2 \Phi_b \tau_2 \right)
 + B{\rm Tr} \left( \Delta \bar \Delta + \Delta^c \bar \Delta^c
\right)
+ {\rm h.c.} \right] \nonumber \\
 &+&  M^2_{ab} {\rm Tr}\left(  \Phi_a^\dag \Phi_b
 \right) + m_0^2(\tilde L^\dagger \tilde L +
 \tilde L^{c\dagger} \tilde L^c) \nonumber \\
 &+& M_\Delta^2 {\rm Tr} \left( \Delta^\dag \Delta + \Delta^{c\dag}
 \Delta^c \right) + M^2_{\bar \Delta} {\rm Tr} \left( \bar \Delta^\dag 
\bar \Delta +
\bar \Delta^{c\dag} \bar \Delta^c \right) \ , \nonumber
\end{eqnarray}
where in addition to the terms dependent on Higgs fields, we have also
kept slepton terms since sneutrino is electrically neutral and can in
principle have vev. We have omitted all terms involving the squarks.
 All soft parameters with mass dimensions are assumed to have TeV
SUSY-breaking scale.

 It was shown in
\cite{kuchi} that if the ground state of this tree level potential has to
break parity, it must break R-parity by giving a vev to the
$\tilde{\nu^c}$ field. We review this
argument in the appendix.

The global minimum of the model is then characterized by the following
vev pattern of the fields
\begin{eqnarray}\label{VEV}
\langle \widetilde L_i^c \rangle = \left( \begin{array}{c}
\langle \widetilde{\nu_i^c}\rangle\\
0
\end{array} \right), \ \langle\Delta^c \rangle = \left( \begin{array}{cc}
0 & 0 \\
v_R & 0
\end{array} \right), \
 \langle\bar \Delta^c\rangle = \left( \begin{array}{cc}
0 & \bar v_R \\
0 & 0
\end{array} \right) \ .
\end{eqnarray}
In this case, there is an upper bound on the $v_R$ scale as discussed in
\cite{kuchi1}, given roughly by
\begin{eqnarray}
\frac{m_0}{f} < v_R < \frac{(A+f\mu_\Delta)}{2f} \ .
\end{eqnarray}
where the parameters $A$ and $m_0$ are susy breaking parameters in
the theory, of the order of a few
hundred GeVs and $f$ is a typical Majorana coupling of the $\Delta$
fields. Using these, we get few TeV upper limit on the right-handed scale
if $f \sim \frac{1}{10}$. A detailed derivation of the upper bound is
given in the appendix.

It has recently been shown that in the case of the model with an additional 
singlet, once one takes one loop corrections into account, in the domain
of parameters where $m^2_{\tilde{l^c}}\leq 0$, R-parity conserving and 
electric charge
conserving minimum indeed becomes the global minimum\cite{babu}. In this
case one generically needs a higher parity breaking scale to get correct
mass spectrum for sleptons. Our discussions in the subsequent sections 
will remain valid for the complementary domain i.e. 
$m^2_{\tilde{l^c}}\geq 0$ in  the presence of loop effects. R-parity conserving ground state
can also be a global minimum if one includes higher dimensional operators
provided the parity breaking scale is above $10^{10}$ GeV\cite{goran}. Since we are interested 
in TeV scale $W_R$, these new operators do not affect our considerations.

In the rest of the paper, we will assume that the component of the right
handed sneutrinos that acquires a vev must align along the electron
flavor direction so that the neutrino masses will remain in the sub-eV
range. It appears that it may be possible to ensure this by choosing the
$A_{ee}$ term associated with $\nu^c_e\nu^c_e\Delta^c$ coupling to be
sufficiently negative while keeping this for other flavors to be
positive. We do not pursue the details of this calculation here.

\section{neutrino mass}

In this section, we address the question of how to understand small
neutrino masses in this model. As is well known, neutrinos acquire Dirac 
masses after electroweak 
symmetry is broken by the $\Phi_{1,2}$ vevs. We assume that only 
$<H^0_u>$ and $<H^0_d>$ fields acquire vevs $\kappa_1$ and $\kappa_2$ 
respectively. In this model, B-L breaking gives large Majorana masses to 
the right handed neutrinos which combined with the the Dirac masses leads 
to the usual type I seesaw mechanism. In the absence of the sneutrino
vev, the type I seesaw mechanism yields sub-eV left-handed neutrino
masses from
TeV-scale right handed-neutrino masses provided the neutrino Dirac Yukawa
couplings $Y_\nu$ are of the same order as that of the electron in the
standard model.
An upper bound on the $v_R$ scale dicted by the dynamics of the model, 
implies that all
elements of the neutrino Yukawa coupling matrix in our model must
have an upper bound of order $10^{-6}$. There are no type II
contributions in the renormalizable supersymmetric left-right model in 
the supersymmetric limit. After SUSY breaking, a finite but small 
type II contribution is induced which can be comparable to the type I 
contribution. This does not affect the analysis of neutrino masses 
done below and we ignore it here. As
we show later, the small Dirac Yukawa couplings needed for understanding
small  neutrino masses
have the important implication, that the gravitino lifetime which is
inversely proportional to $Y^2_\nu$ is long enough that it
can become the dark matter of the Universe.

The presence of right handed sneutrino condensate complicates
the analysis of neutrino masses since it introduces
a mixing term of the form $LH_u$ in the superpotential making the
neutrino-Higgsino-gaugino mass matrix to be a $15\times 15$ mass matrix.
As we show below, if we assume that the right handed sneutrino that
acquires a vev is aligned along the electron flavor direction, one can
still use successive seesaw approximation so that neutrino masses remain
in the sub-eV range as in the simple type I seesaw models prior to
neutrino Higgsino mixing.

To proceed with the neutrino masses, note that the terms in the
superpotential are induced by the sneutrino vev and expand the seesaw
matrix are given by
 \begin{eqnarray}\label{bilinear}
\Delta W = (Y_{\nu})_{ij} x_j L^T_i (i\tau_2)_L H_u + (Y_l)_{ij}
x_j L_i^T (i\tau_2)_L H_u' \ ,
\end{eqnarray}
where $H_u$'s are defined through $\Phi_1 = (H_d', H_u)$ and $\Phi_2 =
(H_d, H_u')$, and
$x_i = \langle \tilde \nu_i^c\rangle$.
These terms mix lepton superfields with the Higgs superfields and
therefore break R-parity. They lead to the mixing of lefthanded neutrinos
 with the neutralinos. These mixing terms enlarge the $6\times 6$ normal seesaw
neutrino mass matrix to $15\times 15$.  However, we can employ
successive decoupling to simplify the analysis of this matrix to get
neutrino masses. For $v_R$ in
the multi-TeV range, the $v_R$-scale fermions decouple first, leaving a
$9\times 9$ mass matrix. On the further application of a second stage
seesaw through integrating out the electroweak scale fields, small
neutrino masses follow.
For convenience of discussion, we choose a basis in which the $Y_l$ is
diagonal and all neutrino mixings arise from $Y_\nu$. We will also
choose all Majorana couplings of right handed neutrinos, $f$, to be
diagonal.

As noted before we will assume the RH-sneutrino vev to align along
the electron flavor i.e. $\langle\tilde{\nu}^c_{\mu,\tau}\rangle=0$ so
that the neutrino-Higgsino mixing is in the MeV range.
The symmetric neutrino-neutralino mass matrix can now be written in the
basis of $\{ \nu_{Li}, \widetilde H_{u}^0, \widetilde H_{d}^0,
\widetilde H_u^{'0}, \widetilde H_d^{'0}, (-i\lambda_{L}^3),
(-i\lambda_R^3), (-i\lambda_{\rm BL}), \nu_{i}^c. \widetilde
\Delta^{0c}, \widetilde{\overline{\Delta}^{0c}} \}$ as
\begin{equation}
\left(
 \begin{array}{ccccccccccc}
0 & (Y_\nu)_{ik} x_k & 0 & (Y_l)_{ik} x_k & 0 & 0 & 0 & 0 &
   (Y_\nu)_{ij} \kappa_1 & 0 & 0 \\
- & 0 & -\mu_{12} & 0 & - \mu_{11} &
   -\frac{g_L}{\sqrt{2}}\kappa_1 & - \frac{g_R}{\sqrt{2}} \kappa_1 & 0 & 0
   & 0 & 0 \\
- & -\mu_{12} & 0 & - \mu_{22} & 0 &
   \frac{g_L}{\sqrt{2}}\kappa_2 & \frac{g_R}{\sqrt{2}} \kappa_2 & 0 & 0 & 0 & 0 \\
- & 0 & -\mu_{22} & 0 & - \mu_{12} & 0 & 0 & 0 & 0
   & 0 & 0 \\
- & -\mu_{11} & 0 & - \mu_{12} & 0 & 0 & 0 & 0 & 0
   & 0 & 0 \\
- & - & - & - & - & M_L & 0 & 0 & 0 & 0 & 0 \\
- & - & - & - & - & 0 & M_R & 0 & \frac{1}{\sqrt{2}} g_R
   x_j & \sqrt{2} g_R v_R & -\sqrt{2} g_R \bar v_R \\
- & - & - & - & - & 0 & 0 & M_{\rm BL} & - \frac{1}{\sqrt{2}}
   g_{\rm BL} x_j & -\sqrt{2} g_{\rm BL} v_R & \sqrt{2} g_{\rm BL} \bar v_R \\
-&- &- &- &- &- &- &- & f_{ij} v_R & f_{ki} x_k & 0 \\
-& -& -& -& -& -&- &- &- & 0 & -\mu_R \\
-& -& -& -& -& -& -& -& -& -\mu_R & 0
\end{array} \right) \ , \nonumber
\end{equation}
where we have assumed the vev's of doublet fields have the following pattern
\begin{eqnarray}
\Phi_1 = \left( \begin{array}{cc}
0 & 0 \\
0 & \kappa_1
\end{array} \right), \ \ \Phi_2 = \left( \begin{array}{cc}
\kappa_2 & 0 \\
0 & 1
\end{array} \right) \ .
\end{eqnarray}
To simplify this matrix to obtain the neutrino masses, we
proceed in two
steps as mentioned: first we integrate out the TeV scale fields
$\nu_i^c$, $\widetilde
\Delta^c$, $\widetilde{\overline{\Delta}^c}$ and a combination of
$\lambda_R^3$, $\lambda_{\rm BL}$. The remaining combination $
\widetilde B = \frac{g_{\rm BL}}{\sqrt{g_R^2 + g_{\rm BL}^2}} (-i\lambda_R^3) +
\frac{g_R}{\sqrt{g_R^2 + g_{\rm BL}^2}} (-i\lambda_{\rm BL})$
stays at the electroweak scale. The mass matrix for the electroweak scale
active fields is
\begin{equation}\label{mixing}
\begin{array}{cl}
& \begin{array}{ccccccc}
           \;\;\;\;\;\;\;\;\;\;\;\;\;\; \nu_{Lj} \;\;\;\;\;\;\;\;\;\;\;\; & \;\;\widetilde H_{u}^0\;\;\;\;\; & \widetilde H_{d}^0\;\;\;\;\; & \widetilde H_u^{'0}\;\;\;\;\; & \widetilde H_d^{'0}\;\;\; & \widetilde W_{3L}\;\;\;\;\;\;\;\;\;\;\;\;\; & \widetilde B
  \end{array} \\
\begin{array}{c}
\nu_{i} \\
\widetilde H_u^0 \\
\widetilde H_d^0 \\
\widetilde H_u^{'0} \\
\widetilde H_d^{'0} \\
\widetilde W_{3L} \\
\widetilde B
\end{array} & \left(
\begin{array}{ccccccc}
- \frac{\kappa_1^2}{v_R} (Y_\nu f^{-1} Y_{\nu}^T)_{ij} &
    (Y_\nu)_{ik} x_k & 0 & (Y_l)_{ik} x_k & 0 & 0 & 0  \\
-& 0 & -\mu_{12} & 0 & - \mu_{11} &
    -\frac{g_L}{\sqrt{2}}\kappa_1 & \frac{g_V}{\sqrt{g_R^2+g_V^2}} \left( -
    \frac{g_R}{\sqrt{2}} \kappa_1 \right)\\
-& -\mu_{12} & 0 & - \mu_{22} & 0 &
    \frac{g_L}{\sqrt{2}}\kappa_2 & \frac{g_V}{\sqrt{g_R^2+g_V^2}} \left(
    \frac{g_R}{\sqrt{2}} \kappa_2 \right) \\
-& 0 & -\mu_{22} & 0 & - \mu_{12} & 0 & 0  \\
-& -\mu_{11} & 0 & - \mu_{12} & 0 & 0 & 0  \\
-&-  &-  &-  &-  & M_L & 0 \\
-&-  &-  &-  &-  & 0 & M_{\widetilde B}
\end{array} \right)
\end{array} \ ,
\end{equation}
where $
M_{\widetilde B} = \frac{g_{\rm BL}^2}{g_{\rm BL}^2 + g_R^2} M_R +
\frac{g_R^2}{g_{\rm BL}^2 + g_R^2} M_{\rm BL} $
and $g'$ satisfies $
g'^{-2} = g_R^{-2} + g_{\rm BL}^{-2}$.

When $Y_{\nu}$ and $Y_l$ are small, we can still use the
seesaw formula to estimate the neutrino mass contributions from R-parity 
violating terms. {\small
\begin{eqnarray}
\delta M_\nu &=& - \left[ (Y_\nu)_{ik} x_k,\  0,\ (Y_l)_{ik} x_k,\
0,\ 0,\ 0 \right] \nonumber \\
&\times& \left[\begin{array}{cccccc}
 0 & -\mu_{12} & 0 & - \mu_{11} & -\frac{g_L}{\sqrt{2}}\kappa_1 & -
\frac{g'}{\sqrt{2}} \kappa_1 \\
 -\mu_{12} & 0 & - \mu_{22} & 0 & \frac{g_L}{\sqrt{2}}\kappa_2 &
\frac{g'}{\sqrt{2}} \kappa_2 \\
0 & -\mu_{22} & 0 & - \mu_{12} & 0 & 0  \\
-\mu_{11} & 0 & - \mu_{12} & 0 & 0 & 0  \\
 -\frac{g_L}{\sqrt{2}}\kappa_1 & \frac{g_L}{\sqrt{2}}\kappa_2  & 0 & 0 &
M_L & 0 \\
- \frac{g'}{\sqrt{2}} \kappa_1  & \frac{g'}{\sqrt{2}} \kappa_2 & 0
& 0 & 0 & M_{\widetilde B}
\end{array} \right]^{-1}\left[\begin{array}{c}
(Y_\nu^T)_{kj} x_k \\
0 \\
(Y_l^T)_{kj} x_k \\
0 \\
0 \\
0
\end{array} \right] \ .
\end{eqnarray} }
The sneutrino condensates picks up 11, 13, 31 and 33 elements of
the inverse matrix.
To apply the seesaw formula,
we concentrate on the corresponding cofactors of these elements.
\begin{eqnarray}
 {\rm cof}_{11} &=& \frac{1}{2} \mu_{12}^2 \kappa_2^2 \left( g_L^2
 M_{\widetilde B} + g'^{2} M_{L} \right) = \frac{ g_L^2 g'^{2}}{2}
 \mu_{12}^2 \kappa_2^2 \left( \frac{M_L}{g_L^2} + \frac{M_R}{g_R^2} +
\frac{M_{\rm BL}}{g_{\rm BL}^2} \right) \ , \nonumber \\
 {\rm cof}_{13} &=& {\rm cof}_{31} = -  \frac{ g_L^2 g'^{2}}{2} \mu_{12}
 \mu_{11} \kappa_2^2 \left( \frac{M_L}{g_L^2} + \frac{M_R}{g_R^2} +
\frac{M_{\rm BL}}{g_{\rm BL}^2} \right) \ , \nonumber \\
{\rm cof}_{33} &=&  \frac{ g_L^2 g'^{2}}{2} \mu_{11}^2 \kappa_2^2
\left( \frac{M_L}{g_L^2} + \frac{M_R}{g_R^2} + \frac{M_{\rm BL}}{g_{\rm BL}^2}
\right) \ .
\end{eqnarray}
Therefore, the neutrino mass can be written as
\begin{eqnarray}\label{numass}
 \left(M_{\nu}\right)_{ij} =&-& \frac{\kappa_1^2}{v_R} (Y_\nu f^{-1}
Y_{\nu}^T)_{ij}
 - \frac{g_L^2 g'^{2} \kappa_2^2}{2 M_L M_{\widetilde B}
( \mu_{11}\mu_{22} - \mu_{12}^2)} \cdot \left(\frac{M_L}{g_L^2} +
\frac{M_R}{g_R^2} + \frac{M_{\rm BL}}{g_{\rm BL}^2}\right) \nonumber \\
&\times& \left[ \left( \frac{\mu_{12}}{\mu_{11}} \right) (Y_\nu)_{ik}
(Y_\nu)_{jk} + x_1^2 (Y_\nu)_{i1} y_e \delta_{j1} + x_1^2 (Y_\nu)_{1j}
y_e \delta_{i1} - \left( \frac{\mu_{11}}{\mu_{12}} \right) y_e^2
\delta_{i1} \delta_{j1} \right].
\end{eqnarray}
where the first term is the usual see-saw formula for neutrino; we have
worked in the basis that the charged lepton Yukawa coupling is diagonal
and assumed only $x_1 \neq 0$.
In the following sections, we will see that this constraint helps us to 
avoid
too large RPV terms, which not only makes gravitino life-time sufficiently
long but also has important collider implications.

An interesting aspect of the neutrino mass formula is the presence of
the term $\left(\frac{M_L}{g_L^2} +
\frac{M_R}{g_R^2} + \frac{M_{\rm BL}}{g_{\rm BL}^2}\right)$. In the
framework of mSUGRA and gauge mediated SUSY
breaking, one typically gets
\begin{equation}
 \frac{|M_L|}{g_L^2} = \frac{|M_{R}|}{g_R^{2}} = \frac{|M_{\rm
BL}|}{g_{\rm BL}^2} \ .
\end{equation}
This relation is preserved by renormalization group evolution. In SUSYLR
case, we have
$M_L=M_R^*$ can be complex, while $M_{\rm BL}$ must  be real. In a
special case where $\arg
(M_L) = \pm 2\pi/3$, the second term in Eq.~(\ref{numass}) vanishes, i.e.
$\delta
M_\nu=0$. In this case, we can also allow $x_2,\ x_3 \neq 0$ without
affecting the neutrino mass discussion above. However no such assumption
is necessary if we only choose $x_1\neq 0$.

Note that in the context of MSSM with R-parity
breaking bilinear terms, getting small neutrino masses
generically requires to fine-tune the neutrino-Higgsino coupling terms
~\cite{valle1,grossman}. In our case, the smallness of these terms is
now guaranteed by the smallness of
the Dirac Yukawa couplings together with the alignment of
$\langle\tilde{\nu}^c\rangle$ along the electron direction.

\section{Sources of R-parity violation in the model}

Our model is different
from all previous models of R-parity breaking \cite{aulakh,valle1}
because R-parity breaking is forced by the tree level dynamics of the
theory to enable parity breaking~\cite{kuchi}: We do not have the freedom
of choosing the tree level parameters leading to an alternative vacuum
with R-parity conservation.
Here R-parity is broken by the right-handed sneutrino vev which
leads to bilinear R-parity violating coupling at low energies. As noted
recently, in models with singlets, once one loop corrections are 
included\cite{babu}, in a
parameter subdomain, one can indeed have an R-parity conserving
vacuum as a global minimum without breaking electric charge. If the same 
discussion were to apply to the minimal case under discussion here,
 the considerations of this paper will still remain valid in the 
complementary parameter domain.

The dominant R-parity violating interaction in the effective
theory below the TeV scale,
 after the right-handed sneutrino and the heavy
$\Delta$ fields are integrated out, is given by
\begin{eqnarray}
\Delta W = \epsilon_i L_i^T(i\tau_2) H_{u} + \epsilon'_1 L_1^T (i\tau_2)
H_{u}' \ .
\end{eqnarray}
There are analogous terms coming from the soft-breaking trilinear terms
given by
\begin{eqnarray}\label{softbilinear}
{\cal L}_1~=~B_i \tilde{L}_i^T(i\tau_2) H_{u} +B'_1\tilde{L}_1^T(i\tau_2)
H_{u}' \ ,
\end{eqnarray}
where $\epsilon_i = (Y_\nu)_{i1} x_1$ and $\epsilon'_i = Y_e x_1 \delta_{i1}$.
The terms in the second equation above arise when we
implement the soft SUSY breaking on the effective R-parity breaking
Lagrangian (say via usual mSUGRA). Since in general the coefficients are
arbitrary and we ignore then in what follows.

In order to understand neutrino masses,
we have assumed that only the right-handed electron sneutrino gets a VEV.
The main consequence of this is that since $Y_e\sim 10^{-5.5}$ and
$Y_{\nu,ij}\leq 10^{-6}$, the strengths of RPV
interactions are all very small. In the presence of these terms, the
gravitino is unstable; however, we will see later that the small
magnitude of these coupling strengths required to understand
neutrino masses, allows the gravitino to live long
enough to be the dark
matter of the Universe. Thus the gravitino dark matter is connected to
the smallness of the neutrino mass, and the viability of the
scenario does not require tuning of a separate parameter.

\begin{figure}[hbt]
\begin{center}
\includegraphics[width=7cm]{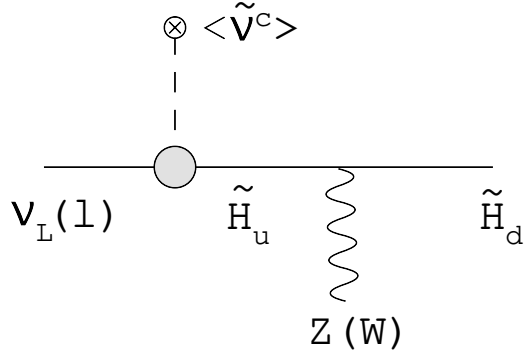}
\caption{``Flavor-violating" gauge couplings due to bilinear
R-parity violation. The shaded blob is an R-Parity breaking vertex.}
\end{center}
\end{figure}

\begin{figure}[hbt]
\begin{center}
\includegraphics[width=7cm]{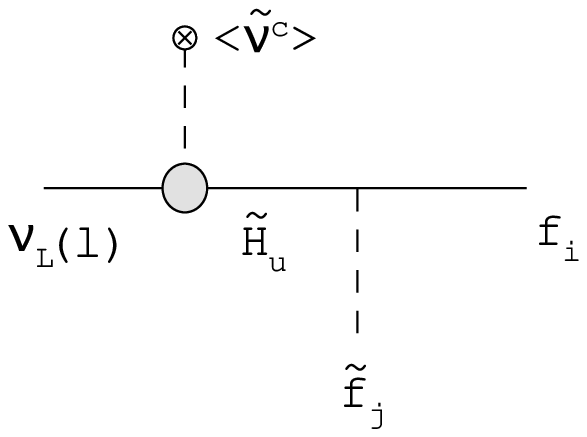}\; \;
\includegraphics[width=7cm]{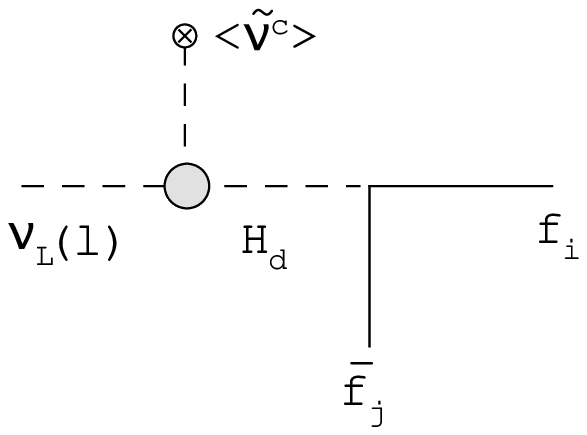}
\caption{The induced tri-linear RPV couplings. The shaded blobs are R-Parity breaking vertices.}
\end{center}
\end{figure}

The bilinear R-parity breaking terms in the superpotential lead to
mixings between neutralinos, $\widetilde H_u$ and neutrinos $\nu$ as
discussed in the
previous section and similarly between charginos and charged leptons.
These terms can also generate mixing between sleptons
and $H_d$ boson through the F-term of $H_u$ and $H_u'$
\begin{eqnarray}\label{mixing2}
F_{H_u} &=& \mu_{11} H_d' + \mu_{12} H_{d} + \epsilon_i L_i \ , \nonumber \\
F_{H_u'} &=& \mu_{21} H_d' + \mu_{22} H_{d} + \epsilon_1' L_1 \ .
\end{eqnarray}
To calculate R-parity violating decays, a general way is
to diagonalize all the mass matrices and
then to rotate all fields into physical states.
In the present work, since we only estimate the order of magnitude for
the R-parity violating effects, we adopt an approximate but more
convenient approach. We
take the bilinear RPV terms as mass perturbations which are to be inserted
into all the amplitudes. It is these mass insertions that act as sources
of R-parity violation. In Fig.~1, we illustrate how to get the
``flavor-violating" gauge coupling induced by
bilinear RPV term~\cite{Roy}, which plays an important role in
neutralino decay. The tri-linear couplings of the conventional
R-parity breaking $\lambda L L E^c$ and $\lambda' Q L D^c$ terms can also be
generated from the bilinear terms. The diagram on the left-panel in Fig.~2
is generated from neutrino-neutralino mixing in Eq.~(\ref{mixing}), and that on
the right is generated through slepton-Higgs mixing due to Eq.~(\ref{mixing2}).
As expected, the right-handed (s)leptons can be coupled to leptons,
whereas the
left-handed (s)leptons couple to both hadrons and leptons.

\section{Gravitino dark matter}

In generic SUGRA theories, the gravitino is a very weakly coupled
particle with mass ranging from eV to many TeV's. It can be produced in
early universe plasma and remains in equilibrium with rest of the cosmic
soup at very high twemperatures i.e. $(T\sim M_{\rm pl})$.
 Slightly once the universe cools below the Planck temperature,
the gravitinos decouple. Since their annihilation or decay rate are very
slow, their number density
dilutes only due to entropy dumped into the cosmic bath at different
annihilation
thresholds of other particles. This dilution is not a large effect.
Therefore in the
absence of inflation, if gravitino is the LSP and R-parity
is conserved, its mass must not to exceed 1 keV in order not to over-close
the universe.

In the inflationary scenario however, any initial gravitino abundance
will be diluted to very tiny values. However, secondary
production of gravitinos in the reheating process can be appreciable
and proportional to the reheat temperature. This has been estimated in
various papers to be ~\cite{grav}.
\begin{eqnarray}
 \Omega_{3/2}h^2~\approx~0.27\left(\frac{T_R}{10^{10}~{\rm
GeV}}\right)\left(\frac{100 ~\rm
GeV}{m_{3/2}}\right)\left(\frac{m_{\tilde{g}}}{1 \rm ~TeV}\right)^2 \ ,
\end{eqnarray}
where $T_R$ is the reheating temperature, and $m_{3/2}$ and $m_{\tilde g}$ are
the gravitino and gluino masses, respectively.
While this would permit gravitinos in the 100 GeV mass range, one must be
mindful of other constraints when R-parity is conserved: the NLSP (often
neutralino or
stau, etc.) decays late, after freeze-out and often after big-bang
nucleosynthesis (BBN) ($T\lesssim$
1~MeV), and produces a large amount of entropy mainly
as photons into the universe, thereby
drastically changing the ratio of $\displaystyle
\frac{n_B}{n_\gamma}$. Also the decay products can
destroy the produced elements making it hard to
understand the successes of BBN~\cite{kohri}. Therefore, in a consistent
picture of the universe described by supergravity theories,
the NLSP must decay quickly
($<10^2$s). If R-pairty is conserved, typical NLSP life time however is
anywhere from a few days to years as can be inferred from the formula
\begin{eqnarray}
\tau_{\rm NLSP} \approx 9~ {\rm days} \left( \frac{m_{3/2}}{100 {\rm
GeV}} \right)^2 \left( \frac{100 {\rm GeV}}{m_{\rm NLSP}}
\right)^5 \ .
\end{eqnarray}
Here we need $m_{3/2} \ll 1$ GeV for $m_{\rm NLSP}$ mass
around 100 GeV.
So whereas the overclosure constraints by gravitinos can be reconciled
with multi-GeV mass gravitinos by adjusting the reheat temperature, the
NLSP lifetime constraints cannot be accommodated  without extreme fine
tuning.

This late-decay problem can be solved if R-parity is violated by a small
amount~\cite{gravilsp}. [For an R-parity conserving theory, there
are also ways to avoid this;  see \cite{howie}.]
In this case, the neutralino decays very quickly into RPV channels and
there is no longer an upper-bound from BBN considerations. Our model
leads naturally
to this scenario since the gravitino can now decay to several channels,
as shown in Fig. 3. To see if it lives long enough to
become a viable dark matter, we need to estimate its lifetime.

\begin{figure}[hbt]
\begin{center}
\includegraphics[width=8cm]{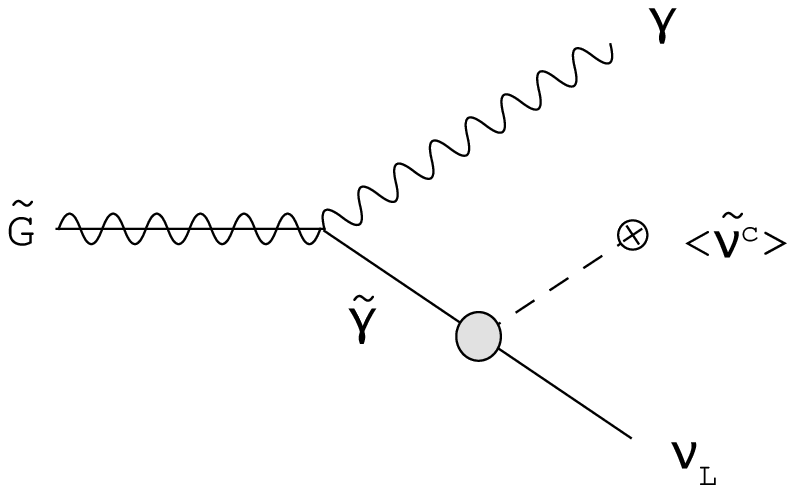}\
\includegraphics[width=6.5cm]{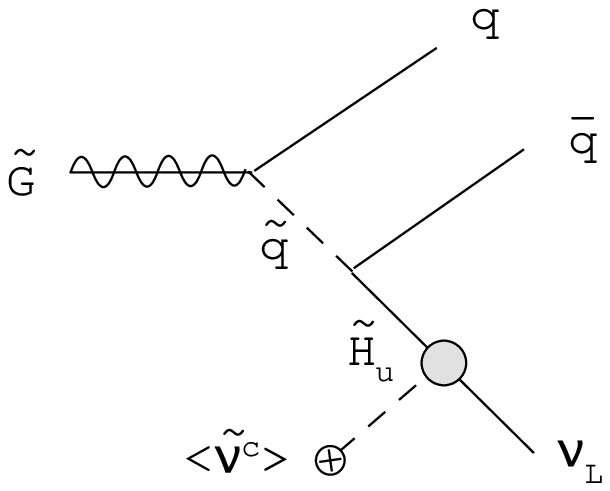}
\caption{Left-panel: Dominant PRV two-body gravitino decay mode  ($\tilde G\rightarrow \gamma \nu_L$) in our model. Right-panel: a typical three-body gravitino decay mode.}
\end{center}
\end{figure}

The dominant decay-modes of gravitino are $\gamma+\nu$,
$W^\pm +l^\mp$, $Z^0+\nu$ and $h^0+\nu$~\cite{nugamma}.
From the following analysis of cosmological constraints, we find the
gravitino should be generally lighter than $W$- or $Z$-boson. In this
case, gravitino mainly decays into photon and neutrino. The strength
of the $\gamma+\nu$ decay that occurs through the diagram depends on
the value of the photino-neutrino mixing. In our model this mixing
occurs via the oneloop diagram in Fig. 4 and the corresponding mixing
parameter $U_{\tilde{\gamma}-\nu}$ can be estimated to be
\begin{eqnarray}
U_{\tilde{\gamma}-\nu}~=~\frac{<\tilde{\nu^c}>v_{wk}sin \beta \mu^2 e
h_e}{16\pi^2 M^3_{susy}M_{\tilde{\gamma}}}\ ,
\end{eqnarray}
and is of order $10^{-6}-10^{-7}$ for $\mu\sim M_{SUSY}\sim
M_{\tilde{\gamma}}\sim <\tilde{\nu^c}> \sim 100$ GeV and $h_e\sim 
10^{-5}$. The decay rate is then given by
\begin{eqnarray}
\Gamma(\tilde G \rightarrow \nu + \gamma) \approx \frac{|U_{\nu
\tilde\gamma}|^2}{32 \pi} \frac{m_{3/2}^3}{M_{\rm
pl}^2} \ ,
\end{eqnarray}
where $|U_{\nu \tilde \gamma}|^2$ is the percentage of photino in
neutrino mass eigenstate in the presence of R-parity violation $ \nu
\approx \nu_0 + U_{\nu \tilde\gamma} \tilde \gamma + \cdots$ as in Fig. 
4.

\begin{figure}[hbt]
\begin{center}
\includegraphics[width=8cm]{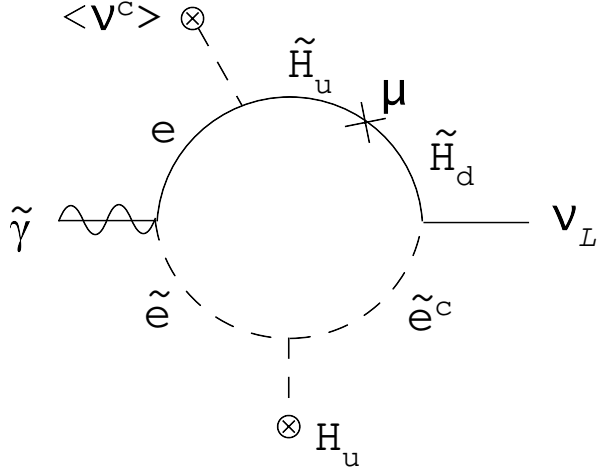}
\caption{1-loop diagram for photino and neutrino mixing.}
\end{center}
\end{figure}

The transverse and longitudinal components of a decaying gravitino give
comparable couplings to final state photon and neutrino. As a rough
estimate, taking $m_{3/2} \sim 2$ GeV
  and $|U_{\nu\tilde
\gamma}|\approx
10^{-6}-10^{-7}$ (as given in our model), one gets $\Gamma(\tilde G 
\rightarrow
\nu + \gamma)^{-1}
\approx 10^{49}-10^{51}$ GeV$^{-1} \approx 10^{25}-10^{27}$ sec.
 Thus we see that even though the gravitino is
unstable via R-parity violating interactions, it can become a viable dark
matter of the Universe.
The gravitino also has three-body decay modes with
$\Gamma_{3-\rm body}^{-1}~=~\left(\frac{h_{\nu}h_q
\langle\tilde{\nu}^c\rangle m^5_{3/2}}{192\pi^3 M_{Pl}M^4_{SUSY}}
\right)^{-1}$ but this rate is small compared to the two body decay rate.


The gravitino decay also produces energetic extra-galatic diffuse
$\gamma$-rays in the universe. From the above decay rate, one can derive
the photon energy-flux per unit solid angle assuming gravitino has
homogeneous distribution throughout the universe~\cite{gravilsp1}
\begin{eqnarray}
\left|E^2 \frac{{\rm d} J (E(t_0))}{{\rm d}E}\right| &=& \frac{\rho_c
\Omega_{3/2} \Gamma(\tilde G \rightarrow \nu + \gamma)}{8 \pi H_0
\Omega_{M}^{1/2}} \left( \frac{2E}{m_{3/2}} \right)^{5/2} \left[1 +
\frac{\Omega_\Lambda}{\Omega_M} \left( \frac{2E}{m_{3/2}}
\right)^{3}\right]^{-1/2} \nonumber \\
&\simeq&  \frac{\rho_c \Omega_{3/2}}{8 \pi \tau_{3/2} H_0
\Omega_{M}^{1/2}} \left( \frac{2E}{m_{3/2}} \right)^{5/2} \left[1
+ \frac{\Omega_\Lambda}{\Omega_M} \left( \frac{2E}{m_{3/2}}
\right)^{3}\right]^{-1/2} \ ,
\end{eqnarray}
where the last step follows if $\tilde G
\rightarrow \nu + \gamma$ is the dominant decay channel of
gravitino. A $\theta$-function $\displaystyle \theta \left( 1
- \frac{2E}{m_{3/2}} \right)$ is implicit because the
photon frequency is cut off at $m_{3/2}/2$, where the flux is
peaked. Experimentally, EGRET observes the extragalatic cosmic
 gamma-ray flux has an excess over the power law spectrum, which can be
as large as $2.23 \times 10^{-6}$ (cm$^2$ str s)$^{-1}$ GeV for photon
energy between 2 and 20
GeV \cite{EGRET}. Attributing part of this excess to
the decay of gravitino dark matter in the universe, one has an
upper bound on the mass of gravitino for a given neutrino-photino mixing
$|U_{\nu \tilde \gamma}|$. Taking the mixing parameter around $10^{-6}$
we find that $m_{3/2} \lesssim 2$ GeV for it to be
consistent with cosmology.

\section{NLSP and Verticed displaced at LHC}

The above R-parity breaking scenario can be tested at LHC. Here we discuss
the possible signatures following from various decays of various
possible NLSP's.

\subsection{Neutralino as NLSP}

{\it R-parity conserving decays}\ \ The next-to-lightest
superparticle can be the neutralino, i.e. a linear
combination of neutral gauginos and higgsinos. It has both RPC and PRV decay
channels. The R-parity conserving channels are $\widetilde \chi^0
\rightarrow
\widetilde G+\gamma, \widetilde G+Z^0, \widetilde G+h^0$, with gravitino
dominantly in the longitudinal component. These
decay rates are calculated in the same way to gravitino decaying into
neutrino and photon~\cite{giu},  yielding
\begin{eqnarray}
\Gamma(\widetilde \chi^0 \rightarrow \widetilde G \gamma) &=&
\frac{|U_{\widetilde \chi^0 \widetilde \gamma}|^2 k^2 m_{\widetilde
\chi^0}^5}{16 \pi F^2} \ , \nonumber\\
\Gamma(\widetilde \chi^0 \rightarrow \widetilde G Z^0) &=&
\frac{|U_{\widetilde \chi^0 \widetilde Z}|^2 k^2 m_{\widetilde
\chi^0}^5}{16 \pi F^2} \left( 1 - \frac{M_{Z^0}^2}{m_{\widetilde
\chi^0}^2} \right)^4 \ ,  \nonumber\\
\Gamma(\widetilde \chi^0 \rightarrow \widetilde G h^0) &=&
\frac{|U_{\widetilde \chi^0 \widetilde h}|^2 k^2 m_{\widetilde
\chi^0}^5}{16 \pi
F^2} \left( 1 - \frac{M_{h^0}^2}{m_{\widetilde \chi^0}^2}
\right)^4 \ ,
\end{eqnarray}
$|U_{\widetilde \chi^0 i}|^2$ is the  percentage of the i-th
species (photino,
zino, higgsino) in the neutralino NLSP and $\frac{k}{F}$ is the coupling
constant of goldstino with matter fields. As will be seen, such decays
are much slower than R-parity violating modes.

\bigskip

\begin{figure}[hbt]
\begin{center}
\includegraphics[width=7.5cm]{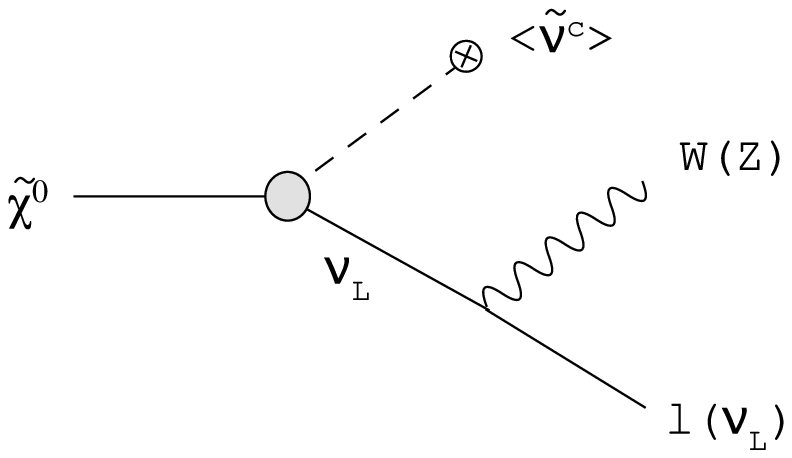}\;
\includegraphics[width=6cm]{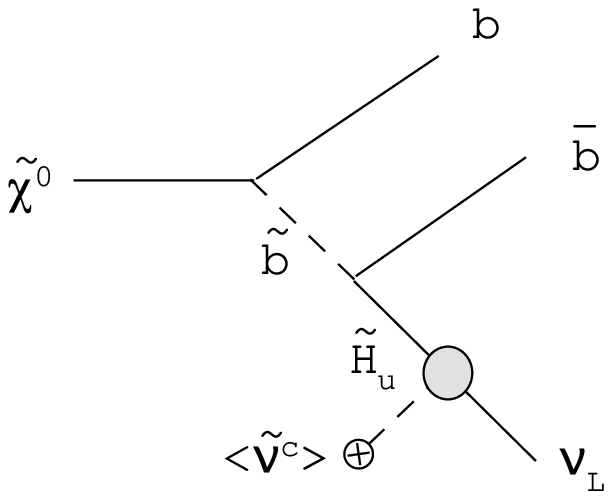}
\caption{A typical two-body (left) and three-body (right) PRV decay diagram of neutralino NLSP.}
\label{neutraldecay}
\end{center}
\end{figure}

{\it R-parity violating two-body decays} The light neutral
gaugino (wino or bino) does not directly couple to right-handed sneutrino
(standard-model charge-free), so the higgsino component in NLSP controls its
R-parity violating decays. If the neutralino NLSP is heavier than the $W$
or $Z$-boson, it can decay into $W^\pm + l^\mp$ or $Z^0 + \nu$. The
corresponding Feynman-diagrams are shown in Fig.~\ref{neutraldecay},
and the decay rates are on the order of
\begin{eqnarray}
\Gamma(\widetilde \chi^0 \rightarrow W^\pm l^\mp) &\approx& \frac{ G_F
m_{\widetilde \chi^0}^3}{8 \sqrt{2} \pi} \left| U_{\widetilde \chi^0
\nu} \right|^2 \left( 1 + \frac{2 M_W^2}{m_{\widetilde \chi^0}^2}
\right) \left( 1 - \frac{M_W^2}{m_{\widetilde \chi^0}^2} \right)^2 \ ,
\nonumber \\
\Gamma(\widetilde \chi^0 \rightarrow Z^0 \nu_L) &\approx& \frac{G_F
m_{\widetilde \chi^0}^3}{32 \sqrt{2}\pi} \left| U_{\widetilde \chi^0
\nu} \right|^2 \left( 1 + \frac{2 M_Z^2}{m_{\widetilde \chi^0}^2}
\right) \left( 1 - \frac{M_Z^2}{m_{\widetilde \chi^0}^2} \right)^2 \ .
\end{eqnarray}

To make a rough estimate, we take $\left| U_{\widetilde \chi^0 \nu}
\right| = 10^{-6}$ and $m_{\widetilde \chi^0} = 100-200$ GeV. We then
find that $\Gamma_{\rm
2-body} \approx10^{-13}- 10^{-12}$ GeV which corresponds to a neutralino
lifetime
of $\sim 10^{-11}-10^{-12}$ sec. This will lead to a vertex displacement
in the
detector of about $0.1-1$ mm.

 This decay rate is fast enough not to
ruin the success of BBN, and slow enough to produce collider signatures
such as vertex displacement at LHC. The displacement of secondary vertex
where NLSP decays is around $c\tau \approx 1$ mm, which is observable
within the detector.

\bigskip

\begin{figure}[hbt]
\begin{center}
\includegraphics[width=8cm]{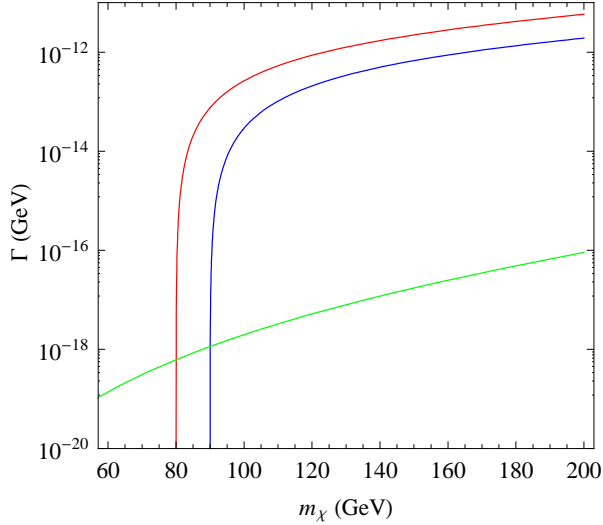}
\caption{Decay rates as a function of neutralino mass. Red curve is for $\widetilde \chi^0 \rightarrow W^{\pm} l^\mp$, blue is for $\widetilde \chi^0 \rightarrow Z^0 \nu$, and green is for $\widetilde \chi^0 \rightarrow \nu_L b_L \bar b_R$, where we choose $|U_{\chi \nu}| \approx 10^{-6}$ and $|U_{\chi \widetilde Z}| \sim O(1)$. We also choose $\tan\beta=3$ and $m_{\widetilde \nu} = 300$ GeV.}
\end{center}
\label{decayrate}
\end{figure}

{\it R-parity violating three-body decay} If the neutralino
is lighter than $W^\pm+l^\mp$ or $Z^0+\nu$ mass, the two-body
RPV decay channels are forbidden. It can only decay into a three-body
final state
through a virtual $W$- or $Z$-boson or sparticles, which couples to
two SM particles. The dominant three-body decay mode is going to  $b+\bar
b+\nu$~\cite{Dreiner:1991pe}, as shown in Fig.~\ref{neutraldecay}. The
three-body decay rates have been calculated in Ref.~\cite{Baltz:1997gd}.
In Fig.~6, we plot the two- and three-body decay rates as a
function of the neutralino NLSP mass. We see that the three-body
decay has a longer lift time and can hardly produce an observable vertex
displacement inside the detector near the vertex.

\subsection{Stau as NLSP}

\begin{figure}[hbt]
\begin{center}
\includegraphics[width=12cm]{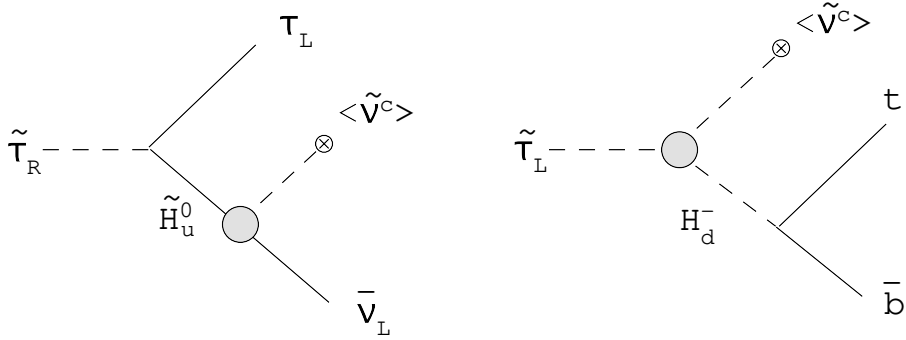}
\caption{Feynman diagrams for stau decay.}
\label{staudecay}
\end{center}
\end{figure}

The stau can also serve as the NLSP in the model. Generally, there is a
small mixing between left- and right-handed staus. The lighter stau
$\widetilde\tau_1$ is a linear combination of them
\begin{eqnarray}
\widetilde \tau_1 = \alpha \widetilde \tau_L + \beta \widetilde \tau_R \ ,
\end{eqnarray}
where $|\alpha|^2 + |\beta|^2=1$.
From the discussions in Section IV, we know that at the R-parity
violating vertices, the right-handed sleptons
mainly couple to lepton final states, while left-handed sleptons can
decay either leptonically or hadronically. Typical diagrams for stau decay
are shown in Fig.~\ref{staudecay}. The decay rates are
\begin{eqnarray}
& &\Gamma(\widetilde\tau_L \rightarrow \bar t b) \approx \frac{3 G_F
m_t^2 m_{\widetilde \tau_1} }{4 \sqrt{2} \pi} \left| U_{\widetilde
\tau_L H^{-} } \right|^2 \left(1 - \frac{m_t^2}{m_{\widetilde \tau_1}^2
} \right)^2 \ , \nonumber \\
& &\Gamma(\widetilde\tau_R \rightarrow \tau \bar \nu) \approx
\frac{G_F m_\tau^2 m_{\widetilde \tau_1}}{4 \sqrt{2} \pi} \left| U_{\nu
\widetilde H^0} \right|^2 \left(1 - \frac{m_\tau^2}{m_{\widetilde
\tau_1}^2} \right)^2 \ ,
\end{eqnarray}
where the mixing due to RPV mass-insertion can be estimated to be
$\left| U_{\widetilde \tau_L H^{-} } \right|, \  \left| U_{\nu
\widetilde H^0} \right| \approx \frac{x_1 Y_{\nu(l)}}{\mu} \approx
10^{-6}$.

\begin{figure}[hbt]
\begin{center}
\includegraphics[width=8cm]{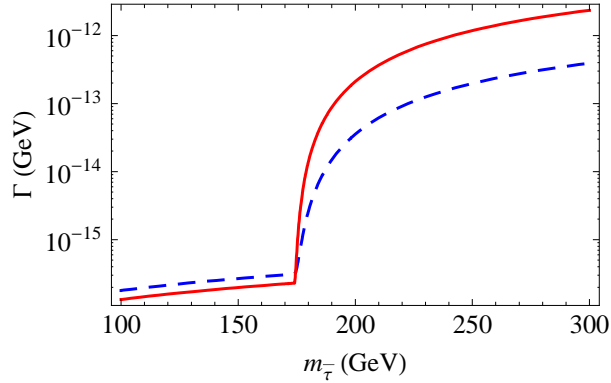}
\caption{Stau NLSP decay rate as a function of its mass. Blue curve is
for $|\alpha|^2=0.05$ and the red one is for $|\alpha|^2=0.3$. There is
a kink at $m_{\widetilde \tau_1} = m_t$ where a significant hadronic
final-state decay channel opens.}
\end{center}
\label{staurate}
\end{figure}

In Fig.~8, we plot the stau decay rate as a function of its
mass. There is a kink at $m_{\widetilde \tau_1} = m_t$ where a
significant hadronic final state decay channel opens.
For stau mass
less than 300 GeV, the decay rate is of the order of
$10^{-13}$ GeV, which corresponds to a vertex displacement 1 mm.

Actually, the NLSP can also be left-handed sneutrino. The decay diagrams
are similar to those of stau NLSP, and one the sneutrino has a
similar decay rate to sleptons.

\section{Additional Comments}
In this section, we discuss some other aspects of the model.

{\bf Strong and SUSY CP} It has been pointed out\cite{rasin} 
that the constraint of left-right symmetry restricts the mass matrices 
and phases in the model in such a way that it provides a solution to both 
the strong CP and SUSY CP problem. Key to solving the strong CP problem 
is the hermiticity of the quark mass matrices in the model. Once R-parity 
is broken, it is not clear that this property will hold. However we find 
that if the right handed sneutrino vev is aligned along the electron 
flavor, the induced phases in the Det $M_q$ are of order 
$(\frac{Y_e<\tilde{\nu^c}>}{\mu})^2\sim 10^{-10}$ which is below the 
bound on the $\theta$ provided by electric dipole  moment of the neutron.
SUSY CP problems are not affected by this phase.

\begin{figure}[hbt]
\begin{center}
\includegraphics[width=6cm]{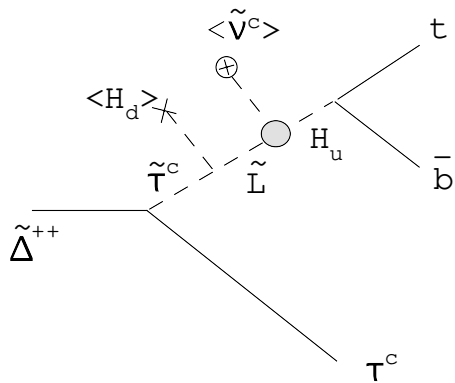}
\caption{A Feynman diagram for doubly-charged Higgsino decay.}
\label{staudecay1}
\end{center}
\end{figure}

{\bf Decays of the doubly-charged Higgsino}
One of the distinguishing features of the B-L=2 triplet Higgses is the 
presence of doubly charged Higgs fields. These are accompanied by their 
fermionic superpartners ($\tilde{\Delta}^{++}$). These particles can 
have masses in the sub-TeV to TeV range and may therefore be accessible 
to LHC. In the presence of R-parity breaking, these fermions will decay 
to $\tilde{\Delta}^{++}\to \tau^c t\bar{b}$ (as shown in Fig.~\ref{staudecay1}). There 
are no standard model 
background for such decay modes.
 
{\bf R-parity breaking processes in the early universe}
We now make a few more comments on the impact of R-parity violating in
our model in the early universe. Note that due to the smallness of the
strength of the $\epsilon_i LH_u$ term in the superpotential i.e. 
$\epsilon_i\sim
10^{-4}$ GeV, the effective $\lambda$ ($LLe^c$) and $\lambda'$ ($QLd^c$)
interactions are of order or less than equal to $10^{-8}$. It was noted
in ref.\cite{sacha}, for such small strengths, the lepton number
violating interactions involving sparticles are out of equilibrium. As a
result one could perhaps contemplate generating lepton asymmetry by the
decays of
the NLSP particle e.g. neutralino slightly above the electroweak scale
and have them converted to baryons via the sphaleron effects. This 
question is under consideration.

\section{Conclusion}

In conclusion, we have discussed the implications of the minimal 
renormalizable supersymmetric left-right seesaw model as an extension of 
MSSM to include neutrino masses and show that the model leads 
to dynamical R-parity violation by its ground state in order to
 break parity symmetry. We
have analyzed a particular realization of the model where
spontaneous R-parity breaking occurs along the $\tilde{\nu^c}_e$
direction. We first show that small neutrino masses in this model can be
understood only by the usual low scale seesaw condition that the Dirac
Yukawa couplings are tuned to the value of $Y_\nu \leq 10^{-6}$ and 
without any further tuning. We then  show that if gravitino is the LSP, 
then its
decays are automatically suppressed by the same condition that guarantees
the smallness of neutrino masses, making the gravitino long lived
enough to be an unstable dark matter of the
Universe. We also point out that the NLSP decays in this model may lead
to displaced vertices which can provide a clear LHC signal.
 These models
have many other collider implications such as doubly charged Higgs and
Higgsino fields\cite{huitu1} that have been discussed extensively in the
literature as well as low energy lepton number violating signals such as
muonium-anti-muonium oscillations\cite{herczeg}.

\bigskip

This work was partially supported by the U. S. Department of Energy
via grant DE-FG02-93ER-40762. R. N. M. is supported by NSF grant
No. PHY-0652363. Y. Z. acknowledges the hospitality and support
from the TQHN group at University of Maryland and a partial support
from NSFC grants 10421503 and 10625521. We thank Y. Grossmann for
comments and reading the manuscript.

\section*{{\large Appendix}}

In this appendix, we discuss the minimization of the Higgs potential to
obtain an upper bound on the right-handed scale $v_R$.
\smallskip

First we will show that in the SUSYLR model, if R-parity is preserved by
the vacuum, parity symmetry cannot be broken i.e. $v_R=\bar{v}_R=0$ as
shown in \cite{kuchi}. A more precise statement perhaps is that, if we
express the potential, $V$ as functions of the vevs of the neutral Higgs
fields ( $v_R, \bar{v}_R$ and $x$ and look for a minimum of $V$ along
the direction $x=0$ (i.e. R-parity conserving ), the minimum occurs at
$v_R=\bar{v}_R=0$. We
then show that once we include R-parity breaking effect by the vacuum
 i.e. $\langle\tilde{\nu^c}\rangle \equiv x \neq 0$ i.e. along the 
direction $x \neq 0$, there
appear global minima that break parity i.e.  $v_R, \bar{v}_R\neq 0$ and
also that it occurs only below a certain value for  $v_R$
and $\bar{v}_R$ i.e. there is an upper limit on the parity breaking
scale. This proves our assertion that R-parity breaking in this theory is
a dynamical phenomenon.

To show this let us start with the potential in Eq.~(2)
which consists of
the field vevs $v_R$ and $\bar v_R$ and $x\equiv \langle\tilde{\nu^c}\rangle$ and
look for its minimum:
\begin{eqnarray}\label{AV}
V &=& \left[M_{\Delta}^2 v_R^2 +
M_{\bar{\Delta}}^2 \bar v_R^2 - 2 |B| v_R \bar v_R
\right] \nonumber \\
&+& \left[ f^2 x^4 -(2Av_R+2f\mu_\Delta \bar{v}_R- m^2_0  - 4 f^2 v^2_R)
x^2
\right] +
\left[ \frac{g_R^2 + g^{2}_{BL}}{8} \left( x^2 - 2 v_R^2 + 2 \bar v_R^2
\right)^2 \right] \ , \nonumber \\
\end{eqnarray}
We have set the $\Phi$ and $\Delta,\bar\Delta$ vevs to zero. Note that we
have kept the $<\tilde{\nu^c}>$ in the potential. For simplicity, we have
set $f~=~f_1$.

The first point to note is that to ensure a lower bound on the potential,
we must satisfy the conditions on the parameters:
\begin{eqnarray}\label{condition}
M^2_\Delta~+~M^2_{\bar{\Delta}}\geq 2 B \ , \nonumber\\
M_\Delta M_{\bar{\Delta}} > B\ . 
\end{eqnarray}
The first constraint comes from looking at the direction
$v_R~=~\bar{v}_R$ and $x=0$ demanding that the potential is bounded from
below. The second comes from looking along the QED breaking vacuum so
that the D-terms vanish and setting $x=0$ and again demanding positivity.
Once these two conditions are imposed, for $x=0$, the minimum of the
potential corresponds to $v_R=\bar{v}_R=~0$ and hence no parity violation.

It is worth pointing out that the form of the potential for $x=0$ is same
as in the case of MSSM, where of course we know that symmetry breaking
occurs. The difference in the case of SUSYLR is the
observation\cite{kuchi} that there exist QED breaking directions along
which for arbitrary $v_R$ and $\bar{v}_R$, the D-term vanishes so that
one has the second condition in Eq. (\ref{condition}). In the case of MSSM, the
second condition does not exist and in fact to break the gauge symmetry
in SUSYLR case,
one needs the opposite of the second condition i.e. $M_\Delta
M_{\bar{\Delta}}< B $.

In order to show that along the nonzero $x$ directions, one can indeed
have a minimum that breaks
the gauge symmetry, it is convenient to rewrite the potential in Eq. (\ref{AV})
as follows:
\begin{eqnarray}
V~=~(f^2+\frac{\tilde{g}^2}{8})[x^2-C]^2+D(v_R, \bar{v}_R) \ ,
\end{eqnarray}
where
\begin{eqnarray}
D(v_R, \bar{v}_R)~&=&~\left[M_{\Delta}^2 v_R^2 +
M_{\bar{\Delta}}^2 \bar v_R^2 - 2 |B| v_R \bar
v_R~+~\frac{\tilde{g}^2}{2}(v^2_R-\bar{v}^2_R)^2
\right]-\left(f^2+\frac{\tilde{g}^2}{8}\right)C^2 \ , \nonumber \\
C~&=&~\frac{\left[2Av_R+2f\mu_\Delta \bar{v}_R- m^2_0  - 4 f^2
v^2_R+\frac{1}{2}\tilde{g}^2(v^2_R-\bar{v}^2_R)\right]}{2(f^2+
\frac{\tilde{g}^2}{8})} \ , 
\end{eqnarray}
where we define $\tilde{g}^2~=~(g^2_R+g^{'2})$.
Advantage of rewriting this way is that we can now minimize with respect
to $x$ very easily and get
\begin{eqnarray}
x^2~=~C.
\end{eqnarray}
Since $x$ is a real number, the above equation implies that $C\geq 0$.
For $v_R$ and $\bar{v}_R$ close to each other, the $C\geq 0$ condition
turns into an upper limit on the $v_R$ scale of
\begin{eqnarray}
 \frac{(A+f\mu_\Delta) -\sqrt{(A+f\mu_\Delta)^2 -4f^2m^2_0}}{4f^2} \leq
v_R \leq
\frac{(A+f\mu_\Delta) + \sqrt{(A+f\mu_\Delta)^2 -4f^2m^2_0}}{4f^2} \ .
\end{eqnarray}
This is clear from the expression for $C$ since for
large $v_R$, the $-f^2v^2_R$ term in $C$ dominates making $C < 0$ and
hence
driving the vev of $x$ to zero in which case, the
minimum corresponds to $v_R~=~\bar{v}_R~=0$ as
noted and hence no parity violation.
This is in accord with the observation of ref.\cite{kuchi1} that the tree
level potential for the minimal SUSYLR model requires R-parity violation
if parity has to break, as required to get the standard model and the
parity breaking scale has an
upper limit in the TeV range, making the theory experimentally testable
at LHC.

 We have done a numerical analysis of the potential and find that
 the full potential indeed has a negative
minimum value only when the $x\neq 0$ and $v_R \neq 0$ and $\bar{v}_R\neq
0$ giving the desired parity violating graound state. We also find
from this numerical analysis that
once we set $C\leq 0$ or equivalently $x~=~0$, the ground state
corresponds to
$v_R~=\bar{v}_R~=0$. In the table below, we give some numerical
examples of solutions for the desired vacua for specific TeV
scale parameters in the potential for the case when $x\neq 0$. We check
that the potential has a
minimum for each case with a value of the potential at the minimum which
is negative so that it is indeed a global minimum.

\begin{table}[hbt]
\begin{tabular}{|c|c|c|c||c|c|c|}
\hline
 $f$ & $A$ (GeV) & $\mu$ (GeV) & $m_0$ (GeV) & $v_R$ (GeV) & $\bar v_R$
(GeV) & $\tilde \nu^c$ (GeV)  \\
\hline
0.1 & 100 & 200 & 500 & 2363 & 1677 & 2286 \\
\hline
0.1 & 100 & 500 & 600 & 2897 & 2166 & 2658 \\
\hline
0.2 & 500 & 200 & 500 & 3561 & 2304 & 4020 \\
\hline
0.3 & 500 & 500 & 500 & 1865 & 1736 & 1815 \\
\hline
\end{tabular}
\caption[]{ The other parameters are chosen to be $M_{\Delta} = 1$ TeV,
$M_{\bar \Delta} = 1.1$ TeV and $B = 1.099$ TeV$^2$.}
\end{table}

Incidentally, along the $x=0$ direction in the potential, there is also
no electrweak symmetry breaking by similar arguments as above
\cite{kuchi} i.e. $\langle\phi\rangle=0$; however along the above $x\neq 0$
direction, one immediately gets $\langle\phi\rangle\neq 0$ also along with breaking
of $SU(2)_R\times U(1)_{B-L}$ symmetry.



\end{document}